\documentclass{PoS}

\usepackage{amssymb} 
\usepackage{amsmath} 
\usepackage{bm}

\title{
Six-dimensional regularization of chiral gauge theories on a lattice\footnote{
This is a joint contribution for two talks 
``Six-dimensional regularization of chiral gauge theories on a lattice (I \& II)'' given by two of the authors.
}}

\ShortTitle{Six-dimensional regularization of chiral gauge theories}

\author{
        \speaker{Hidenori~Fukaya},
        Tetsuya~Onogi, 
        Shota~Yamamoto, 
        and Ryo~Yamamura$^\dagger$
        \\
        \\
        \\
        \llap{$$}
        Department of Physics, Osaka University, 
        Toyonaka, Osaka 560-0043 Japan
}

\abstract{
We propose a six-dimensional regularization of four dimensional chiral gauge theories. 
We consider a massive Dirac fermion in six dimensions with two different operators
having domain-wall profiles in the fifth and the sixth directions,
respectively. A Weyl fermion appears as a localized mode at the
junction of the two domain-walls. In our formulation, the Stora-Zumino
chain of the anomaly descent equations, starting from the axial $U(1)$ anomaly in six-dimensions to the gauge
anomaly in four-dimensions, is naturally embedded. Moreover, a similar
inflow of the global anomalies is found. The anomaly free condition is
equivalent to requiring that the axial $U(1)$ anomaly and the parity
anomaly are canceled among the six-dimensional Dirac fermions. Putting
the gauge field at the four- dimensional junction and extending it to
the bulk using the Yang-Mills gradient flow, as recently proposed by
Grabowska and Kaplan, we define the four-dimensional path integral of
the target chiral gauge theory.
}

\FullConference{34th annual International Symposium on Lattice Field Theory\\
		24-30 July 2016\\
		University of Southampton, UK}

\begin{document}

\section{Introduction}

In our textbooks on quantun field theory \cite{Weinberg},
we learn that the consistent form of
gauge anomaly of the Weyl fermions is  
obtained from the axial $U(1)$ anomaly in 6-dimensions 
via 5-dimensional parity anomaly:
\begin{eqnarray}
\text{Tr}[F^3] &=& 
d \text{Tr}\left[ A(dA)^2 +\frac{3}{2}A^3dA + \frac{3}{5}A^5\right],\\
&&\delta_v \text{Tr}\left[ A(dA)^2 +\frac{3}{2}A^3dA + \frac{3}{5}A^5\right] 
= d \left\{ v^a\textcolor{black}{\text{Tr}\left[ T^a d (AdA +\frac{1}{2}A^3)\right]}\right\}.
\end{eqnarray}  
Here we have used the differential form for the gauge field $A=A_\mu dx^\mu$,
and $F=dA+A^2$ with the external derivative $d$, 
and gauge transformation function $v=v^aT^a$ with the 
generators $T^a$ of $SU(N)$ gauge group.
This is the so-called Stora-Zumino chain of the anomaly descent equations
\cite{Stora:1983ct, Zumino:1983ew, Zumino:1983rz},
and the obtained consistent gauge anomaly is known to be
the unique solution to the Wess-Zumino condition \cite{Wess:1971yu}.
The 5-th and 6-th dimensions in Refs.\cite{AlvarezGaume:1983cs, Sumitani:1984ed} 
are introduced to describe a 2-parameter family of continuous interpolation from trivial
gauge field to the reference gauge field and the gauge transformations.
However, it is not clear whether these extra-dimensions have physical importance
beyond a mathematical tool to derive the consistent anomaly.

In this work \cite{Fukaya:2016ofi}, we try to give more physical meaning of the
extra-dimensions in a formulation of Weyl fermions.
Namely, we consider the 5-th and 6-th directions as
real coordinates, and put a massive Dirac fermion 
in  6-dimensional Euclidean space-time.
Since this starting point is a massive and vector-like
fermion system, we expect that a naive lattice discretization by 
the Wilson fermion is sufficient 
to non-perturbatively regularize it.

A Weyl fermion appears as a low-energy state localized
on a 4-dimensional junction in the 6-dimensions.
This situation is realized by putting a part of
the 6-dimensional bulk space in a ``topological'' phase.
More concretely, we introduce two kinds of domain-wall masses,
perpendicular to each other, which give a gap
to the 6-dimensional fermion in the bulk, but
produces a gapless mode at the junction of the domain-walls.
We may consider this set-up as a ``doubly gapped topological
insulator'' which has a gapless mode at ``the edge of the edges''.

We find  that
the Stora-Zumino anomaly chain is naturally embedded in our 6-dimensional Dirac fermion determinant.
The total determinant can be decomposed into 6-dimensional bulk,
and 5, and 4-dimensional edge localized mode's contributions.
The complex phase of the 6-dimensional bulk modes in the topological phase
is nothing but the axial $U(1)$ anomaly.
Since this $U(1)$ anomaly is a total divergence, its parity violation
appear as a 5-dimensional surface contribution, which is compensated by
the parity anomaly appearing in the phase of the 5-dimensional modes.
The gauge transformation of the parity anomaly of the 5-dimensional
modes is again a total divergence and its 4-dimensional surface
contribution is then canceled by the gauge anomaly of the Weyl fermion
at the 4-dimensional domain-wall junction.

Since the axial $U(1)$ symmetry is a global symmetry, 
its anomaly causes no theoretical problem in constructing a gauge theory in 6-dimensions.
It is, however, an obstruction to give a 4-dimensional local
effective theory, since
it contains the information of whole 6-dimensional gauge configuration
even in the low-energy limit.
If we are forced to consider such a theory as one in the 4-dimensional
effective theory, it is no longer a local field theory.
The anomaly-free sets of fermion flavors are the special
exceptions where we can have a local 4-dimensional effective action.
For the anomaly free fermion contents, 
the gauge and parity anomalies are canceled among flavors and never flow into the 6-dimensional bulk.
 
The two domain-walls must carry different quantum numbers. 
If they were constructed by the same operator,
for example, the ordinary mass term in 6-dimensions,
the resultant 4-dimensional mode would be not chiral but
vector-like fermion having both of left- and right-handed modes.
Therefore, it is a non-trivial issue how to choose the two 
domain-wall operators.
We find several combinations of the two domain-walls producing
the Weyl mode at the junction, and in fact, what we chose for
the presentation at the conference turned out to be
not appropriate for having non-local cancellation of the
physical 4-dimensional phase of the determinant\footnote{
We thank D.~B.~Kaplan for pointing out this problem.
}.
Here, we take another option: a combination of a simple mass term
and an axial vector in 6-th direction to give the domain-wall profile.

Since the axial vector operator is invariant
under the axial $U(1)$ rotation, 
our second domain-wall: a kink structure
of the axial vector operator,
has nothing to do with the Stora-Zumino anomaly ladder.
It is, however, not invariant under another anomalous symmetry:
it flips the sign under reflection of 5-th coordinate.
Since this symmetry is identical to the one yielding
the parity anomaly in 5-dimensions,
we call it $P'$ symmetry (which is different from ordinary 
parity $P$ symmetry in 6-dimension).
To consider physical importance of $P'$ symmetry,
let us consider a single fundamental Dirac fermion
in $SU(2)$ gauge theory.
This fermion is known to have no axial $U(1)$ anomaly,
which results in no perturbative gauge anomaly in 4-dimensions.
However, this fermion can still have non-trivial zero-modes
related to the mod-two index theorem reflecting
the non-trivial homotopy group $\pi_5(SU(2))=\mathbb{Z}_2$.
Now let us ``assume'' that this mod-two index is precisely
counted by $P'$ transformation. Then our second domain-wall
should mediate its effects from the 5-dimensional bulk
to the 5- and 4-dimensional edges,
which may be regarded as the origin 
of global anomaly \cite{Witten:1982fp}.
In fact, we find a ladder of mod-two type indices:
\begin{eqnarray}
\pi_5(SU(2))=\mathbb{Z}_2 
\;\;\;\to\;\;\; \pi_4(SU(2))=\mathbb{Z}_2 \;\;\;\to\;\;\; 
\mbox{complex phase of 4-d Weyl fermion}.
\end{eqnarray}
Thus, our model contains an interesting possibility of
the global anomaly as the consequence of 
the mod-two instantons in the 6-dimensions.

This work is strongly motivated by the recent work by 
Grabowska and Kaplan \cite{Grabowska:2015qpk},
who proposed to construct the chiral gauge theory
using the Yang-Mills gradient flow in the 5-th direction
of their domain-wall fermion formulation,
to maintain the exact 4-dimensional gauge invariance.
We find that their idea can be easily extended 
to our 6-th dimension and we use the gradient flow
to define the 4-dimensional
path integration of the chiral gauge theory.
In fact, taking the first domain-wall mass to infinity,
our model converges to their 5-dimensional model.
In this sense, our 6-dimensional proposal 
may be an extension of their work to incorporate
the global anomaly.
However, we need a further investigation 
to confirm this possibility since the use of the 
Yang-Mills gradient flow makes the role of global anomalies 
obscure: it seems not to allow 
non-trivial mod-two index in 6-dimensional bulk.

The rest of this article is organized as follows.
In Sec.~\ref{sec:model}, we define our model
and show how a Weyl fermion appears at the domain-wall junction.
Then in Sec.~\ref{sec:StoraZuminoladder}
we show that the Stora-Zumino anomaly chain is
naturally embedded in our 6-dimensional Dirac fermion determinant.
In Sec.\ref{sec:globalanomaly} we discuss how 
the global anomalies appear in our model.
The anomaly free condition in our model can be
reduced to the cancellation of the complex phase
in the 6-dimensional bulk, which comes from
the global anomalous symmetries of 
the 6-dimensional Dirac fermions (Sec.~\ref{sec:anomalyfree}).
Finally we propose a lattice regularization in Sec.~\ref{sec:lattice}
and give a summary and discussion in Sec.~\ref{sec:summary}.


\section{Our 6-dimensional model}
\label{sec:model}
In this work, we consider a 6-dimensional Euclidean space-time
whose boundary condition for both of the fermions and bosons 
is periodic in every direction.
Since we are interested in the 4-dimensional gauge theory, 
we simply take the 5-th and 6-th components of the gauge fields to be zero, {\it i.e.}
$A_5(x)=A_6(x)=0$. For the other components, we require
$A_\mu (\mu=1,2,3,4)$ to be symmetric under $x_5\to -x_5$ and $x_6\to -x_6$
(later we give more concrete form of $x_5$- and $x_6$-dependences of $A_\mu$ by the Yang-Mills gradient flow).

Our 6-dimensional determinant of
a Dirac fermion with the Pauli-Villars field is defined by
\begin{eqnarray}
\exp(-W_{\rm 2DW})\equiv \det \left(\frac{D^{\rm 6D}+M\epsilon(x_6)
+i\mu\epsilon(x_5)\gamma_6\gamma_7  }{D^{\rm 6D}+M+i\mu\gamma_6\gamma_7  }\right),
\label{eq:detDW}
\end{eqnarray}
where $D^{\rm 6D}$ denotes the 6-dimensional Dirac operator with a gauge group $SU(N_c)$,
$\epsilon(x)=x/|x|$ denotes the sign function,
and $\gamma_i$ are the $8\times 8$ gamma matrices,
with which the chiral operator is defined as $\gamma_7=i\prod_{i=1}^6\gamma_i$.
Note that the two domain-walls parametrized by $M$ (we call the $M$ domain-wall) and $\mu$ ($\mu$ domain-wall)
intersect with each other at $x_5=x_6=0$, where a Weyl fermion appears.
Since the fermion fields satisfy periodic boundary conditions,
there also exist anti-domain-walls in the determinant.
Although the anti-domain-walls do not appear in the expressions,
we always assume that they are there, and 
will explicitly write them whenever it is necessary.

$M$ denotes the conventional Dirac fermion mass, 
whose domain-wall structure is known to produce massless fermions on it.
As we will see below, this mass term is sensitive to 
the axial $U(1)$ transformation and is the origin of the perturbative
gauge anomaly, which appears at the junction of the two domain-walls.

The $\mu$ term or the axial vector in the 6-th direction
\begin{eqnarray}
i\mu\bar{\psi}\gamma_6\gamma_7\psi,
\end{eqnarray}
on the other hand, is insensitive to the axial $U(1)$ rotation but 
flips its sign under a ``parity'' transformation
$P'$ or the reflection with respect to the $x_5$ axis:
\begin{eqnarray}
P'\psi(x) = i\gamma_5 R_5 \psi(x),\;\;\;\bar{\psi}(x)P'=iR_5\bar{\psi}(x)\gamma_5,
\end{eqnarray}
where $R_i$ denotes the reflection of the $i$-th coordinate: $R_i f(x_i)=f(-x_i)$.
Note that this parity is different from the conventional parity:
\begin{eqnarray}
P\psi(x)=\gamma_1\prod_{i\neq 1}R_i\psi(x),\;\;\;\bar{\psi}(x)P=\prod_{i\neq 1}R_i\bar{\psi}(x)\gamma_1,
\end{eqnarray}
where we take $i=1$ to be the temporal direction.
The main difference is that $P^{'2}=-1$ while $P^2=1$.

The $P'$ symmetry has an anomaly \cite{Redlich:1983dv, Niemi:1983rq, AlvarezGaume:1984nf}.
Since $\{D^{\rm 6D},P'\}=0$, every eigenvalue $i\lambda$ 
of $D^{6D}$ has its pair $-i\lambda$, except for the zero modes.
Therefore, under $P'$, the massless fermion action is manifestly invariant,
while the zero-mode's contribution to the fermion measure Jacobian
is not, since $P'$ flips its sign,
\begin{eqnarray}
D\bar{\psi_0}P'DP'\psi_0 = - D\bar{\psi_0}D\psi_0.
\end{eqnarray}
Note that those from non-zero modes always cancel with their partners.
Therefore, the $P'$ transformation counts the number of zero-modes $\mathfrak{I}$.
Here we would like to stress that $\mathfrak{I}$ can be different from 
the conventional topological charge, or the axial $U(1)$ index $\mathcal{P}$.
An example is the $SU(2)$ theory where the $U(1)_A$ anomaly is exactly zero.
Nevertheless, there exists the so-called mod-two index 
related to the homotopy group $\pi_5(SU(2))=\mathbb{Z}_2$.
This homotopy group happens to be isomorphic to $\pi_4(SU(2))=\mathbb{Z}_2$,
which describes the global anomaly \cite{Witten:1982fp}.
It is then natural to ``assume'' that the number of instantons are given by
$\mathfrak{I}=\mathcal{P}+\mathcal{I}$, where $\mathcal{P}$ controls the perturbative gauge anomaly,
while $\mathcal{I}$ is the origin of global anomalies.

As we have set $A_5(x)=A_6(x)=0$, we have $D^{\rm 6D} = D^{\rm 4D} + \gamma_5\partial_5 + \gamma_6\partial_6$.
Then we can easily find a solution of the Dirac equation
\begin{eqnarray}
\label{eq:4Dsolution}
(D^{\rm 6D}+M\epsilon(x_6)+i\mu\epsilon(x_5)\gamma_6\gamma_7  )\psi(x)=0,
\end{eqnarray}
localized at the domain-wall junction $x_5=x_6=0$ as
\begin{eqnarray}
\psi(x)&=&e^{-M|x_6|}e^{-\mu|x_5|}\phi(\bar{x}),\\
D^{\rm 4D} \phi(\bar{x})&=&0,\\
\label{eq:cond1}\gamma_6\phi(\bar{x})&=&\phi(\bar{x}),\\
\label{eq:cond2}i\gamma_5\gamma_6\gamma_7 \phi(\bar{x})&=&\phi(\bar{x}),
\end{eqnarray}
where $\bar{x}=(x_1,x_2,x_3,x_4)$ and we have assumed $M>0$ and $\mu>0$.
Note that the latter two conditions Eqs.(\ref{eq:cond1}) and (\ref{eq:cond2}) on $\phi(\bar{x})$
require the field to have a positive chirality
(the opposite chirality can be realized by flipping the sign of $M$ and $\mu$).
Namely, we have a single Weyl fermion localized at the domain-wall junction.

As a final remark of this section, we point out that
the choice of the second domain-wall operator is not unique\footnote{
A proposal of using the pseudoscalar operator for the second domain-wall
was given by Neuberger in Ref.~\cite{Neuberger:2003yg}.
We also discuss other possibilities in our paper \cite{Fukaya:2016ofi}.
}.
Here we have chosen the axial vector but
other operators could reproduce the Weyl fermion mode at the junction as well.
As mentioned in the introduction, some of them fail to 
reproduce the physical complex phase of the 4-dimensional effective theory.
We need further investigation how to choose
 the appropriate combinations of the two domain-walls.

\section{Stora-Zumino anomaly ladder}
\label{sec:StoraZuminoladder}

Now let us look into the structure of the anomaly ladder.
This can be performed by decomposing the total Dirac fermion determinant
into 6-, 5- and 4-dimensional mode's contributions.
Details are shown in our paper \cite{Fukaya:2016ofi}.

Let us begin with a simpler set-up taking $\mu=0$:
\begin{eqnarray}
\label{eq:mu0limit}
\det \left(\frac{D^{\rm 6D}+M\epsilon(x_6)}{D^{\rm 6D}+M}\right).
\end{eqnarray}
We may regard this fermion determinant as a model for
the 6-dimensional topological insulator, whose boundary is set at $x_6=0$ plane.
This determinant is real and therefore, its complex phase can be written as 
$i\pi\mathfrak{I}$.
Now we insert determinants with an intermediate cut-off $M_2$, 
to numerators and denominators (thus canceling each other)
and take the $M\gg M_2\gg 0$ limit:
\begin{eqnarray}
\label{eq:mu0limit}
\det \left(\frac{D^{\rm 6D}+M\epsilon(x_6)}{D^{\rm 6D}+M}\right)
&=&
\det \left(\frac{D^{\rm 6D}+M\epsilon(x_6)+iM_2\gamma_6\gamma_7  }{D^{\rm 6D}+M}\right)
\nonumber\\
&&\times\det \left(\frac{D^{\rm 6D}+M\epsilon(x_6)}{D^{\rm 6D}+M\epsilon(x_6)+iM_2\gamma_6\gamma_7  }\right).
\end{eqnarray}
Here the first determinant corresponds to the 6-dimensional bulk contribution 
with an IR cut-off $M_2$ and the second one is the 5-dimensional edge mode's contribution
whose UV cut-off is given by $M_2$.

We find by the Fujikawa's method, that the first determinant in Eq.(\ref{eq:mu0limit})
produces the axial $U(1)_A$ anomaly:
\begin{eqnarray}
\label{eq:U(1)anomaly}
{\rm Im}\ln \det \left(\frac{D^{\rm 6D}+M\epsilon(x_6)+iM_2\gamma_6\gamma_7 }{D^{\rm 6D}+M 
}\right)
&=& \pi\int d^6 x \frac{1-\epsilon(x_6)}{2}\frac{1}{6(4\pi)^3}
\epsilon^{\mu_1\cdots\mu_6}{\rm tr}[F_{\mu_1\mu_2}F_{\mu_3\mu_4}F_{\mu_5\mu_6}]
\nonumber\\
&=& \pi \mathcal{P}^{6D}_{x_6<0}+\pi CS,
\end{eqnarray}
where $ \mathcal{P}^{6D}_{x_6<0}$ counts the bulk instanton number 
in the region $x_6<0$, and $CS$ is the Chern-Simons term on the $M$ domain-wall,
\begin{eqnarray}
CS&\equiv& -\int_{x_6=0} d^5 x \frac{2}{3(4\pi)^3}
\epsilon^{\mu_1\cdots\mu_5}{\rm tr}\left[\frac{1}{2}A_{\mu_1}F_{\mu_2\mu_3}F_{\mu_4\mu_5}
-\frac{i}{2}A_{\mu_1}A_{\mu_2}A_{\mu_3}F_{\mu_4\mu_5}
-\frac{1}{5}A_{\mu_1}A_{\mu_2}A_{\mu_3}A_{\mu_4}A_{\mu_5}
\right].\nonumber\\
\end{eqnarray}

In the second determinant of Eq.(\ref{eq:mu0limit}), only the boundary localized mode 
satisfying $\gamma_6\psi=\psi$ and $(\gamma_6\partial_6+M\epsilon(x_6))\psi=0$, at the $M$ domain-wall can contribute. 
Rearranging the gamma-matrices, one obtains
\begin{equation}
\lim_{M\to \infty}\det \left(\frac{D^{\rm 6D}+M\epsilon(x_6)}{D^{\rm 6D}+M\epsilon(x_6)+iM_2\gamma_6\gamma_7 }\right)
= \det \left(\frac{\bar{D}^{\rm 5D}}{\bar{D}^{\rm 5D}+M_2}\right),
\label{eq:5Ddet}
\end{equation}
where the determinant in the RHS is taken in the reduced space 
of  4$\times$4 gamma matrices $\bar{\gamma}_i$, and 
the corresponding Dirac operator is given by
 $\bar{D}^{\rm 5D}=\sum_{i=1}^5\bar{\gamma}_i'\nabla_i|_{x_6=0}$,
where $\bar{\gamma}_i'=i\bar{\gamma}_5\bar{\gamma}_i$.

It is known that the complex phase of the 5-dimensional 
massless Dirac fermion determinant in Eq.~(\ref{eq:5Ddet}) is
given by the so-called $\eta$-invariant 
\cite{AlvarezGaume:1985di, DellaPietra:1986qd, Kaplan:1995pe}, 
we have obtained a mathematical formula
\begin{eqnarray}
\mathfrak{I} = \mathcal{P}^{6D}_{x_6<0}+CS - \frac{\eta^{\rm 5D}}{2},
\end{eqnarray}
known as the Atiyah-Patodi-Singer index theorem 
\cite{Atiyah:1975jf,Atiyah:1976jg,Atiyah:1980jh}.

Now let us take $\mu$ finite, and take the limit $M\gg \mu\gg 0$:
\begin{eqnarray}
\label{eq:MDW}
\det \left(\frac{D^{\rm 6D}+M\epsilon(x_6)+i\mu\epsilon(x_5) \gamma_6\gamma_7  }{D^{\rm 6D}+M+i\mu\gamma_6\gamma_7  }\right)
&=&   
\det \left(\frac{D^{\rm 6D}+M\epsilon(x_6)+i\mu\gamma_6\gamma_7 }{D^{\rm 6D}+M+i\mu\gamma_6\gamma_7 }\right)
\nonumber\\
&&\hspace{-0.25in}\times\det \left(\frac{D^{\rm 6D}+M\epsilon(x_6)+i\mu\epsilon(x_5)\gamma_6\gamma_7 }
{D^{\rm 6D}+M\epsilon(x_6)+i\mu\gamma_6\gamma_7 }\right).
\end{eqnarray}
The first determinant in Eq.(\ref{eq:MDW}) gives the same contribution as 
the one in Eq.~(\ref{eq:mu0limit}),
{\it i.e.} they yield the same contribution $\pi(\mathcal{P}^{6D}_{x_6<0}+CS)$
to the phase of the determinant.
This is consistent with the axial $U(1)_A$ insensitivity of the $\mu$ domain-wall.

The second determinant in Eq.(\ref{eq:MDW}) in the $M\to \infty$ limit, becomes
\begin{eqnarray}
\det\left(\frac{\bar{D}^{\rm 5D}+\mu\epsilon(x_5)}{\bar{D}^{\rm 5D}+\mu}\right),
\label{eq:det5DonMDW}
\end{eqnarray}
which corresponds to the standard 5-dimensional domain-wall fermion determinant
\cite{Grabowska:2015qpk}.

As shown in \cite{Grabowska:2015qpk}, or in our paper taking 
the anti-domain-wall effects more explicitly, the complex phase $-i\pi\eta^{\rm 5D}/2$
(let us keep this notation given in the $\mu=0$ case) of
the determinant Eq.~(\ref{eq:det5DonMDW}) can be decomposed as
\begin{eqnarray}
\label{eq:ladderMDW}
-\frac{1}{2}\eta^{\rm 5D}&=& -CS^{(x_5<0)} - \mathcal{I}^\text{5D}_{M\gg\mu}- \frac{1}{2}\eta^{\rm 4D}+\frac{\phi^{anom}}{\pi},
\end{eqnarray}
where we have another $CS$ term restricted to the $x_5<0$ region \cite{Callan:1984sa}: 
\begin{eqnarray}
-\pi CS^{(x_5<0)} &\equiv& \pi\int_{x_6=0} d^5 x \frac{4}{3(4\pi)^3}\frac{1-\epsilon(x_5)\epsilon(L_5-x_5)}{2}
\epsilon^{\mu_1\cdots\mu_5}{\rm tr}\left[\frac{1}{2}A_{\mu_1}F_{\mu_2\mu_3}F_{\mu_4\mu_5}
\right.\nonumber\\&&\left.
-\frac{i}{2}A_{\mu_1}A_{\mu_2}A_{\mu_3}F_{\mu_4\mu_5}
-\frac{1}{5}A_{\mu_1}A_{\mu_2}A_{\mu_3}A_{\mu_4}A_{\mu_5}
\right],
\end{eqnarray}
and $\mathcal{I}^\text{5D}_{M\gg\mu}$ counts the number of exotic instantons in 5-dimension.
As 4-dimensional contributions from the Weyl fermion, we have a gauge invariant part denoted by $\eta^{\rm 4D}$,
and the anomalous part $\phi^{anom}$ whose gauge symmetry breaking exactly 
cancels that in $CS^{(x_5<0)}$.

More explicitly, we obtain in the $\mu=\infty$ limit,
\begin{eqnarray}
\lim_{\mu\to \infty}\det\left(\frac{\bar{D}^{\rm 5D}+\mu\epsilon(x_5)}{\bar{D}^{\rm 5D}+\mu}\right)=
\exp\left(-i\pi CS^{(x_5<0)}- i\pi\mathcal{I}^\text{5D}_{M\gg\mu}\right)\times \lim_{\mu_2\to \infty}\det\frac{\mathcal{D}}{\mathcal{D}+\mu_2},
\label{eq:4ddet}
\end{eqnarray}
where $\mathcal{D}$ is 
\begin{eqnarray}
\mathcal{D} = P^5_-\bar{D}^{\rm 4D}P^5_+ + P^5_+\bar{\partial}^{\rm 4D}P^5_-,
\end{eqnarray}
with $\bar{D}^{\rm 4D}=\sum_{i=1}^4\bar{\gamma}'_i\nabla_i|_{x_6=x_5=0}$ and $P^5_\pm=(1\pm \bar{\gamma}_5)/2$.
As will be shown later, we define the bulk gauge fields from  the 4-dimensional gauge fields
at the junction in such a way that the Dirac operator
$\bar{\partial}^{\rm 4D}=\sum_{i=1}^4\bar{\gamma}'_i\nabla_i|_{x_6=0,x_5=L_5}$ becomes 
that for a (almost) free fermion, so that the negative chirality mode
at $x_5=L_5$ is decoupled from the theory.

What we have obtained is the anomaly ladder 
\begin{eqnarray}
\label{eq:ladderMDW}
\phi^\text{total}/\pi
&=& \mathcal{P}^{6D}_{x_6<0}+CS - \frac{\eta^{\rm 5D}}{2},\nonumber\\
\frac{1}{2}\eta^{\rm 5D}&=& CS^{(x_5<0)} + \mathcal{I}^\text{5D}_{M\gg\mu}+ \frac{1}{2}\eta^{\rm 4D}-\frac{\phi^{anom}}{\pi},
\end{eqnarray}
where $\mathcal{P}^{6D}_{x_6<0}$ denotes the 6-dimensional $U(1)_A$ anomaly,
$CS$ and $CS^{(x_5<0)}$ represent the 5-dimensional parity anomaly,
and $\phi^{anom}$ is the source of the consistent gauge anomaly.
This result is consistent with the anomaly descent equations 
found by Stora \cite{Stora:1983ct} 
and Zumino \cite{Zumino:1983ew, Zumino:1983rz},
including the overall constant determined by 
Alvarez-Gaum\'e and Ginsparg \cite{AlvarezGaume:1983cs} 
and Sumitani \cite{Sumitani:1984ed}.

\section{Global anomaly ladder?}
\label{sec:globalanomaly}

In order to trace the anomaly inflow via the $\mu$ domain-wall,
let us take the limit $\mu\gg M\gg 0$.
As will be shown below, the $\mu$ term is insensitive to
the $U(1)_A$ rotation and there is no anomaly inflow
related to the axial $U(1)_A$ anomaly, and therefore,
the conventional perturbative gauge anomaly.
A natural question is then to ask if there is any other 
anomaly flowing into the $\mu$ domain-wall or not.
The answer is yes.
As pointed out in Sec.~\ref{sec:model}, 
we know an example of the exotic instantons and in 
$SU(2)$ gauge theory, which are insensitive to the $U(1)_A$ rotation.
We assume here that these exotic instantons can be precisely
detected by the $P'$ anomaly.
Then, we should have a non-trivial anomaly inflow 
through the $\mu$ domain-wall because it is 
not invariant under $P'$ transformation.

Similarly to the previous section, our goal of this section 
is to decompose the complex phase of the determinant into three parts:
\begin{eqnarray}
\label{eq:muDW}
\text{Im}\ln \det \left(\frac{D^{\rm 6D}+M\epsilon(x_6)+i\mu\epsilon(x_5) \gamma_6\gamma_7  }{D^{\rm 6D}+M+i\mu\gamma_6\gamma_7  }\right)
&=& \phi^\text{total} = \phi^\text{6D} + \phi^\text{5D} + \phi^\text{4D},
\end{eqnarray}
in the limit of  $\mu\gg M\gg 0$.

First, let us decompose the determinant as
\begin{eqnarray}
\label{eq:muDW}
\det \left(\frac{D^{\rm 6D}+M\epsilon(x_6)+i\mu\epsilon(x_5) \gamma_6\gamma_7  }{D^{\rm 6D}+M+i\mu\gamma_6\gamma_7  }\right)
&=& 
\det \left(\frac{D^{\rm 6D}+i\mu\epsilon(x_5)\gamma_6\gamma_7 +M}{D^{\rm 6D}+i\mu\gamma_6\gamma_7 +M}\right)
\nonumber\\
&&\times\det \left(\frac{D^{\rm 6D}+i\mu\epsilon(x_5)\gamma_6\gamma_7 +M\epsilon(x_6)}{D^{\rm 6D}+i\mu\epsilon(x_5)\gamma_6\gamma_7 +M}\right),
\end{eqnarray}
where the first part corresponds to the 6-dimensional bulk contribution and
the second is the one from 5- and 4-dimensional boundary.

Note again that unlike the $M$ domain-wall, the first determinant of Eq.~(\ref{eq:muDW}) does not produce the axial $U(1)$ anomaly. 
Due to the explicit violation of the $SO(6)$ Lorentz symmetry by the axial vector background,
the phase $\phi^\text{6D}$ of the second determinant can be expanded in an $SO(5)$ invariant series of $1/\mu$,
except for the non-perturbative zero mode's contribution 
$\pi \mathcal{I}^{\rm 6D}_{x_5<0}$, which is located in the region $x_5<0$.
More explicitly, we obtain by one-loop computation 
\begin{eqnarray}
\phi^\text{6D} = \pi \mathcal{I}^{\rm 6D}_{x_5<0} + \mu \phi^{(1)} + \mathcal{O}(1/\mu),
\end{eqnarray}
where, the leading order contribution has a form of the Chern-Simons term
\begin{eqnarray}
 \phi^{(1)}=c_0\pi\int d^6 x \frac{4}{3(4\pi)^3}\frac{1-\epsilon(x_5)}{2}
\epsilon^{i_1\cdots i_5}{\rm tr}\left[\frac{1}{2}A_{i_1}F_{i_2 i_3}F_{i_4 i_5}
-\frac{i}{2}A_{i_1}A_{i_2}A_{i_3}F_{i_4 i_5}
-\frac{1}{5}A_{i_1}A_{i_2}A_{i_3}A_{i_4}A_{i_5}
\right],\nonumber\\
\end{eqnarray}
where $c_0$ is a constant.
Since $\phi^{(1)}$ cancels among the anomaly free fermion contents,
 the only non-trivial phase is given by
$\pi \mathcal{I}^{\rm 6D}_{x_5<0}$ when the perturbative gauge anomaly is absent.

The second determinant of Eq.~(\ref{eq:muDW}) in the $\mu\to \infty$ limit converges to\footnote{
As $\hat{D}^{\rm 5D}$ and $P_+$ do not commute with $M\epsilon(x_6)$, we make the order of
the matrix operations explicit.} 
\begin{eqnarray}
\label{eq:4thdetmuDW}
&&\det \left(P_+(\hat{D}^{\rm 5D}+M)^{-1}(\hat{D}^{\rm 5D}+M\epsilon(x_6))P_+ +P_-\right)
\nonumber\\&&
\times\det \left(P_+(\hat{D}^{\rm 5D}-M)^{-1}(\hat{D}^{\rm 5D}-M\epsilon(x_6))P_+ +P_- \right),
\end{eqnarray}
where $\hat{D}^{\rm 5D}=(\sum_{i=1}^4\bar{\gamma}_i \nabla_i+\bar{\gamma_5}\partial_6)|_{x_5=0}$,
and $P_\pm \equiv (1\pm\bar{\gamma}_5)/2$ are Hermitian projection operators.
Unlike the case in the previous section, what we obtain here on the $\mu$ domain-wall is not
a single Dirac fermion but two (4-component) Dirac fermions having Pauli-Villars 
masses $\pm M$ with opposite signs, 
that are constrained to have the positive eigenvalue of the gamma matrix $\bar{\gamma}_5$.
This expression is {\it almost} real, except for the domain-wall $x_6=0$,
since the complex phase comes from the non-commutativity of $\hat{D}^{\rm 5D}$
and $M\epsilon(x_6)$, which is proportional to $\delta(x_6)$.
We can thus express the phases of the 5- and 4-dimensional contributions as
\begin{eqnarray}
\phi^\text{5D} &=& \pi \mathcal{I}^{\rm 5D}_{x_6<0},\\
\phi^\text{4D} &=&-\frac{\pi}{2}\eta^{\rm 4D}, 
\end{eqnarray}
assuming that the $M\gg\mu\gg 0$ and $\mu\gg M\gg 0$ limits commute
to obtain the same gauge invariant part of $\phi^\text{4D}$ as in the previous section. 
It is natural to assume that $\mathcal{I}^{\rm 5D}_{x_6<0}$ gives the 
mod-two instanton found by Witten \cite{Witten:1982fp} which describes the global gauge anomaly.

Note in Eq.~(\ref{eq:4thdetmuDW}), only the Weyl fermion with positive $M$ 
appears in the low-energy limit, while the other leaves a non-local phase,
which cannot be described by any local action.
We consider the latter non-local phase as the contribution of the Chern-Simons,
which automatically cancels the perturbative gauge anomalies.

To summarize this section,
we have confirmed the anomaly inflow via $\mu$ domain-wall,
\begin{eqnarray}
\phi^\text{total} = \pi \mathcal{I}^{\rm 6D}_{x_5<0} +
\pi \mathcal{I}^{\rm 5D}_{x_6<0}-\frac{\pi}{2}\eta^{\rm 4D} + \mu \phi^{(1)} +\mathcal{O}(1/\mu),
\end{eqnarray}
where the mod-two type indices $\mathcal{I}^{\rm 6D}_{x_5<0}$ and $\mathcal{I}^{\rm 5D}_{x_6<0}$ are
balanced in a non-trivial way with the 4-dimensional phase $\eta^{\rm 4D}$ 
(even when the perturbative gauge anomaly and $\phi^{(1)}$ are absent).


\section{Anomaly free condition}
\label{sec:anomalyfree}

In the previous two sections, we have traced two different anomaly inflows taking $M\gg \mu \gg 0$ 
and $\mu \gg M \gg 0$ limits.
At finite $M$ and $\mu$, the situation can be more complicated but
the non-trivial cancellation of anomalies among different dimensions
should be maintained to keep the gauge invariance of the total theory.
In the end, a single Weyl fermion always appears at the junction of the two domain-walls.

\begin{figure}[htbp]
\begin{center}
\includegraphics[width=10cm, angle=90]{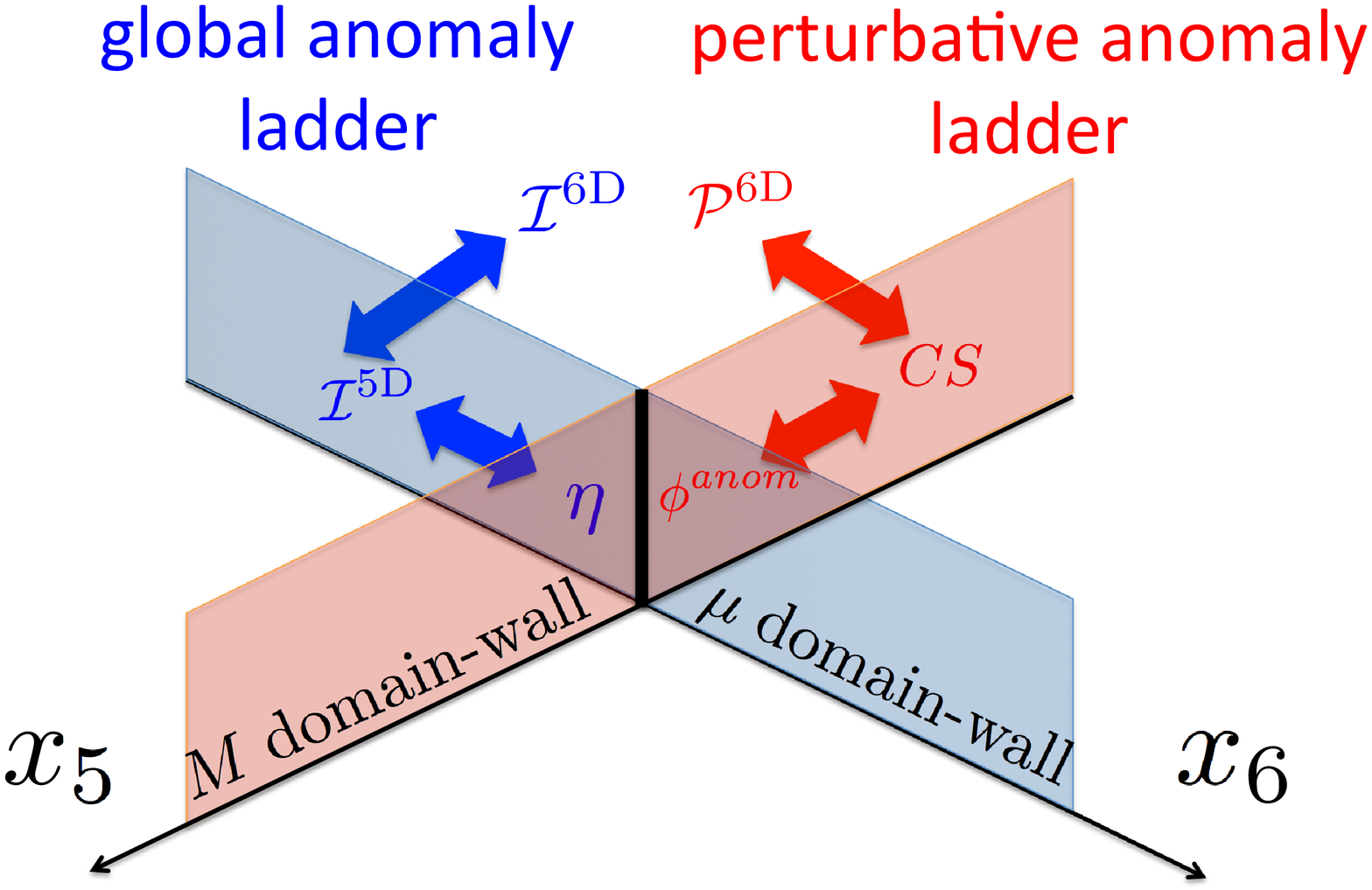}
\caption{The anomaly inflows through the two domain-walls. 
The $M$ domain-wall at $x_6=0$ 
mediates the perturbative anomaly inflow (red arrows), 
which exhibits the Stora-Zumino descent equations.
The $\mu$ domain-wall at $x_5=0$ mediates 
the inflow of the global anomaly (blue). 
} 
\label{fig:2DW}
\end{center}
\end{figure}

When a small gauge transformation is performed at the 4-dimensional junction,
the gauge current flows through the $M$ domain-wall,
but never flows into the $\mu$ domain-wall, 
since there is no CS term which can absorb the gauge non-invariance.
Instead, a large gauge transformation can create exotic instantons on
the $\mu$ domain-wall and flip the sign of the partition function. 
Thus, we confirm that the perturbative anomaly inflow, 
which naturally exhibits the Stora-Zumino anomaly descent equations,
is mediated by the $M$ domain-wall, while the inflow of the global anomaly 
goes through the $\mu$ domain-wall 
(see Fig.~\ref{fig:2DW}).

Now the perturbative and global gauge anomalies are viewed
as these missing gauge current inflows into the 5- and 6-dimensional bulk.
If these currents happen to cancel among different flavors 
(of the 6-dimensional Dirac fermion),
the heavy bulk mode fermions can be decoupled from the low-energy theory,
leaving the 4-dimensional effective theory of Weyl fermions.
Thus, the anomaly free condition is to require the cancellation of 
the $U(1)_A$ anomaly and the $P'$ anomaly in 6-dimensional Dirac fermions.
This condition is equivalent to the cancellation of the phase of 
the bulk fermion determinant mod $2\pi\times$ integer, 
which agrees with the recent discussion by Witten \cite{Witten:2015aba}.
Note that the 4-dimensional edge modes at the domain-wall junction 
are still allowed to have their own complex phase.

The cancellation of the $U(1)_A$ anomaly is guaranteed if
\begin{eqnarray}
\label{eq:pfree}
\sum_L {\rm tr}T_L^a\{T_L^b,T_L^c\}
-\sum_R {\rm tr}T_R^a\{T_R^b,T_R^c\}=0,
\end{eqnarray}
where $T_{L/R}$ denote the gauge group generators
in $L/R$ representation of the corresponding 
left/right handed fermions.
This condition assures the cancellation of 
the $U(1)_A$ anomaly, as well as the CS term
on the $M$ domain-wall, so that the gauge current
never flows out of the 4-dimensional junction.

The cancellation of the global anomalies 
is more non-trivial, as discussed in \cite{Dai:1994kq, Witten:2015aba}.
The global anomaly should be absent not 
only on a simple manifold like $S^4$ or $S^5$
but also on any compact manifold. 
Our set-up on the 6-dimensional torus having
domain-wall junctions of 4-dimensional torus,
is already such a non-trivial example.

Here we take the most conservative condition:
\begin{eqnarray}
\label{eq:gfree}
\mbox{number fermions in the fundamental representation = even},
\end{eqnarray}
after the irreducible decomposition. 
Since the phase from the the exotic index $\mathcal{I}$ is
always multiplied by $2\pi\times$ integer, the global anomaly effect is absent
in the 5- and 6-dimensional bulkmodes.
Note that the standard model of particle physics satisfies the above condition if 
we identify $e/6$ as a unit charge of the hyper-charge.

\section{Lattice regularization}
\label{sec:lattice}

Since our 6-dimensional formulation is based on 
a massive Dirac fermion, it is natural to assume that
we can non-perturbatively 
regularize it on a lattice using the Wilson Dirac operator.

First let us put our lattice in a 6-dimensional
finite box. In particular, we take the 5-th and 6-th coordinates
in the ranges $-L_5 < x_5\leq L_5$ and $-L_6 < x_6\leq L_6$
and assume the periodic boundary condition of the
Dirac fields in every direction.
Because of the boundary condition, we need
 (at least) one $M$ anti-domain-wall at $x_6=L_6 (=-L_6)$
and one $\mu$ anti-domain-wall at $x_5=L_5 (=-L_5)$.
Then we have 4 domain-wall junctions.
Two Weyl fermion modes with positive chirality appear 
at $(x_5,x_6)=(0,0)$ and $(0,L_6)$, while
those with negative chirality are localized at
$(x_5,x_6)=(L_5,0)$ and $(L_5,L_6)$.

Among these 4 junctions, only the one at $(x_5,x_6)=(0,0)$
is needed to formulate the Weyl fermion in 4-dimensions.
Therefore, the other massless fermions 
at other three junctions have to be decoupled from the total theory.
To achieve this, we follow the idea in \cite{Grabowska:2015qpk}
using the Yang-Mills gradient flow in the fifth and sixth directions.
The gradient flow exponentially weaken the gauge fields with the flow time
so that the Weyl fermions at $x_5=L_5$ and $x_6=L_6$ are decoupled from the gauge fields\footnote{
Recently it was proved that this is 
true except for those coming from the non-trivial topologies. 
See Refs.\cite{Okumura:2016dsr,Makino:2016auf,Grabowska:2016bis}
}. 
As flowed fields transform in the same way as the original fields under the gauge transformation,
we can maintain the 4-dimensional gauge invariance of the total theory.

More concretely, we pick up a set of link variables $\{U_\mu(\bar{x})\}(\mu=1,\cdots 4)$
on the 4-dimensional junction at $(x_5,x_6)=(0,0)$.
Then we solve the lattice version of the Yang-Mills gradient flow equation,
\begin{eqnarray}
\frac{\partial}{\partial t}U^t_\mu(\bar{x})=-\left\{\partial_{x,\mu}S_G(U^t)\right\}U^t_\mu(\bar{x}),
\end{eqnarray}
using $U^0_\mu(\bar{x})=U_\mu(\bar{x})$ as the initial condition, 
where $\partial_{x,\mu}S_G(U^t)$ denotes the Lie derivative of
the gauge action $S_G(U^t)$ with respect to $U^t_\mu(\bar{x})$, to define
\begin{equation}
U_\mu(\bar{x},x_5,x_6)=U^{|x_5|+|x_6|}_\mu(\bar{x}).
\end{equation}
Here we always set $U_5=U_6=$unity.
Note that the resulting link variables $U_\mu(\bar{x},x_5,x_6)$ are 
symmetric under $x_5\to -x_5$ and $x_6\to -x_6$.

\begin{figure}[htbp]
\begin{center}
\includegraphics[width=10cm, angle=90]{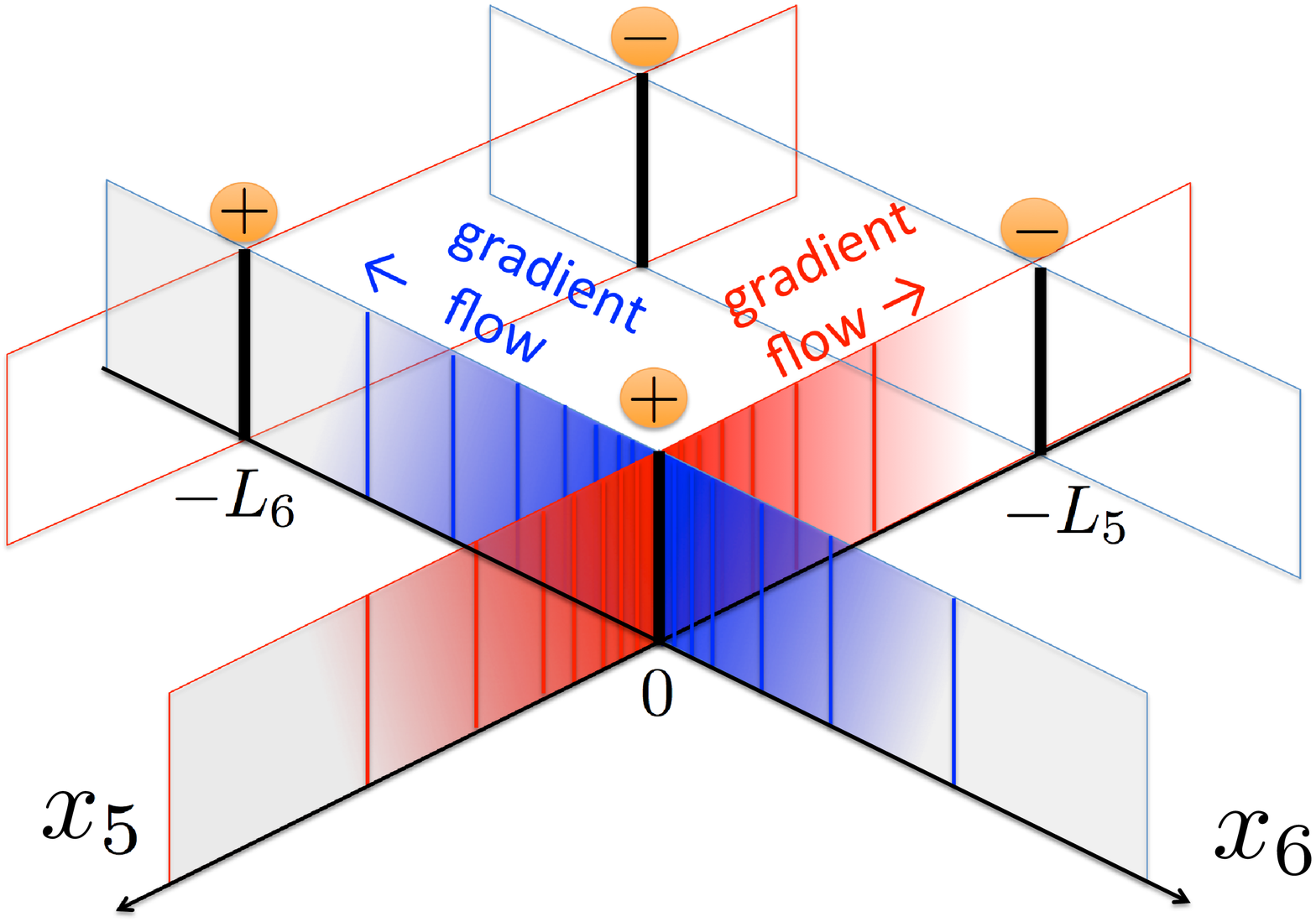}
\caption{Schematic view of our 6-dimensional finite lattice.
The $\pm$ symbols show the Weyl modes with positive and negative chiralities,
localized at each of the four domain-wall junctions.
The case with $M>0$ and $\mu>0$ is shown.
Our target Weyl fermion with positive chirality is localized at
the origin, while other three Weyl fermions are decoupled from
the gauge fields by the gradient flow.
} 
\label{fig:FiniteLattice}
\end{center}
\end{figure}

Finally we ``define'' the 4-dimensional path integral of
anomaly free theory with Weyl fermions.
Together with the gauge part of the action $S_G(\{U_\mu(\bar{x})\})$,
we define
\begin{eqnarray}
\int DU_\mu(\bar{x}) e^{-S_G(\{U_\mu(\bar{x})\})}\prod_{i} \exp \left[-W_{\rm lat}^i(\{U_\mu(\bar{x})\})\right],
\end{eqnarray}
where 
\begin{eqnarray}
\exp \left[-W_{\rm lat}^i(\{U_\mu(\bar{x})\})\right]=\hspace{2in}\nonumber\\
\det \left(\frac{D_W^{{\rm 6D}R_i}+M_i\epsilon(x_6-a/2)\epsilon(L_6-x_6-a/2)
+i\mu_i\epsilon(x_5-a/2)\epsilon(L_5-x_5-a/2)
\gamma_6\gamma_7 }{D_W^{{\rm 6D}R_i}+M_i+i\mu_i\gamma_6\gamma_7 }\right),
\nonumber\\
\end{eqnarray}
where $D_W^{{\rm 6D}R_i}$ denotes the Wilson Dirac operator
in the $R_i$ representation of the gauge group, and 
$M_i$ and $\mu_i$ are chosen to be positive/negative for positive/negative chiral modes.
Note that the Wilson term has to have an opposite sign to $M_i$ and $\mu_i$.
These mass parameters are to be of the order of the lattice cut-off $1/a$.
However, to avoid contamination from the doubler modes, $M_i$ and $\mu_i$  
should satisfy some upper bounds.
We always assume that the set of fermion flavors satisfy the anomaly free conditions
Eqs.~(\ref{eq:pfree}) and (\ref{eq:gfree}).
Our set-up is presented in Fig.~\ref{fig:FiniteLattice}.



\section{Summary and discussion}
\label{sec:summary}

We have proposed a regularization of
the chiral gauge theories in 4-dimensions,
using massive 6-dimensional Dirac fermions.
Using the two different kinds of domain-walls, 
we have succeeded in localizing a single Weyl fermion at the
junction of the domain-walls.
One domain-wall is made giving
a kink mass in the 6-th direction to the fermions,
while another domain-wall is made by giving a kink structure in the 5-th direction 
to a background operator which is insensitive to the $U(1)_A$ rotation.
Our set-up can be viewed as a ``doubly'' gapped topological insulator,
whose 4-dimensional edge modes become massless. 

The domain-wall of the conventional mass term
mediates the perturbative anomaly inflow
and naturally exhibits the chain of the 6-dimensional $U(1)_A$, 
5-dimensional parity, and 4-dimensional gauge anomalies,
known as the descent equations
found by Stora \cite{Stora:1983ct} 
and Zumino \cite{Zumino:1983ew, Zumino:1983rz}.
On another domain-wall, the fermions are forced to form
(almost) a real representation and only mediates the mod-two type anomaly,
which we have assumed to be the source of the global anomalies.

The anomaly free condition of the target 4-dimensional gauge theory
is translated to the cancellation of the axial $U(1)$ and $P'$ anomalies
for a set of 6-dimensional Dirac fermions.
This condition removes the total complex phase from the bulk part of 
the fermion determinant, while the 4-dimensional edge modes can have their own phase.
Using the Yang-Mills gradient flow in the 5-th and 6-th directions,
we can control the remnant of the gauge non-invariance due to the finite cut-offs,
and decouple the Weyl fermions at the junctions of anti-domain-walls.
As our formulation is nothing but a massive vector-like theory,
we expect that a non-perturbative regularization on a lattice is possible,
using standard Wilson Dirac fermions.

There are still a lot of open issues to be investigated.
There is an arbitrariness in the choice of the $\mu$ domain-wall operator,
to realize a single Weyl fermion at the domain-wall junction.
It is also unclear if the $\mu$ domain-wall and associated $P'$ anomaly 
necessarily and sufficiently classify the global anomalies.

It is an interesting question if our formulation can be
extended to a model with physical 
extra dimensions also in the gauge sector.
Such a direction may be linked to studies of higher dimensional 
beyond the standard models.
Our formulation suggests that there is a possibility
of doubly gapped topological insulators in four-dimensions, 
having a conducting mode on two-dimensional edges,
which may be realized in condensed matter systems.
Finally, it would be great if we can incorporate the Higgs field 
to our 6-dimensional lattice and give
{\it a non-perturbative definition} of the standard model,
which is also an interesting subject for further study.

We thank S.~Aoki, D.~Grabowska, K.~Hashimoto, D.~B.~Kaplan, Y.~Kikukawa, H.~Suzuki, and S.~Yamaguchi for useful discussions.
This work is supported in part by the Grand-in-Aid of the Japanese Ministry of Education 
Nos. 25800147, 26247043 (H.F.), No. 26400248(T.O.), and No. 15J01081 (R.Y.).


\begin{thebibliography}{99}
\bibitem{Weinberg}
S.~Weinberg, ``The Quantum Theory of Fields II,'' Cambridge University Press (1995)



\bibitem{Stora:1983ct} 
  R.~Stora,
  LAPP-TH-94, C83-09-01.

\bibitem{Zumino:1983ew} 
  B.~Zumino,
  In *Treiman, S.b. ( Ed.) Et Al.: Current Algebra and Anomalies*, 361-391 and Lawrence Berkeley Lab. - LBL-16747 (83,REC.OCT.) 46p

\bibitem{Zumino:1983rz} 
  B.~Zumino, Y.~S.~Wu and A.~Zee,
  Nucl.\ Phys.\ B {\bf 239}, 477 (1984).
  doi:10.1016/0550-3213(84)90259-1


\bibitem{Wess:1971yu} 
  J.~Wess and B.~Zumino,
  Phys.\ Lett.\  {\bf 37B}, 95 (1971).
  doi:10.1016/0370-2693(71)90582-X




\bibitem{AlvarezGaume:1983cs} 
  L.~Alvarez-Gaume and P.~H.~Ginsparg,
  Nucl.\ Phys.\ B {\bf 243}, 449 (1984).
  doi:10.1016/0550-3213(84)90487-5

\bibitem{Sumitani:1984ed} 
  T.~Sumitani,
  J.\ Phys.\ A {\bf 17}, L811 (1984).
  doi:10.1088/0305-4470/17/14/016


\bibitem{Fukaya:2016ofi} 
  H.~Fukaya, T.~Onogi, S.~Yamamoto and R.~Yamamura,
  arXiv:1607.06174 [hep-th].

\bibitem{Witten:1982fp} 
  E.~Witten,
  Phys.\ Lett.\ B {\bf 117}, 324 (1982).
  doi:10.1016/0370-2693(82)90728-6

\bibitem{Grabowska:2015qpk} 
  D.~M.~Grabowska and D.~B.~Kaplan,
  Phys.\ Rev.\ Lett.\  {\bf 116}, no. 21, 211602 (2016)
  doi:10.1103/PhysRevLett.116.211602
  [arXiv:1511.03649 [hep-lat]].




\bibitem{Redlich:1983dv} 
  A.~N.~Redlich,
  Phys.\ Rev.\ D {\bf 29}, 2366 (1984).
  doi:10.1103/PhysRevD.29.2366

\bibitem{Niemi:1983rq} 
  A.~J.~Niemi and G.~W.~Semenoff,
  Phys.\ Rev.\ Lett.\  {\bf 51}, 2077 (1983).
  doi:10.1103/PhysRevLett.51.2077

\bibitem{AlvarezGaume:1984nf} 
  L.~Alvarez-Gaume, S.~Della Pietra and G.~W.~Moore,
  Annals Phys.\  {\bf 163}, 288 (1985).
  doi:10.1016/0003-4916(85)90383-5


\bibitem{Neuberger:2003yg} 
  H.~Neuberger,
  hep-lat/0303009.



\bibitem{AlvarezGaume:1985di} 
  L.~Alvarez-Gaume, S.~Della Pietra and V.~Della Pietra,
  Phys.\ Lett.\ B {\bf 166}, 177 (1986).
  doi:10.1016/0370-2693(86)91373-0

\bibitem{DellaPietra:1986qd} 
  S.~Della Pietra, V.~Della Pietra and L.~Alvarez-Gaume,
  Commun.\ Math.\ Phys.\  {\bf 109}, 691 (1987).
  doi:10.1007/BF01208963

\bibitem{Kaplan:1995pe} 
  D.~B.~Kaplan and M.~Schmaltz,
  Phys.\ Lett.\ B {\bf 368}, 44 (1996)
  doi:10.1016/0370-2693(95)01485-3
  [hep-th/9510197].


\bibitem{Atiyah:1975jf} 
  M.~F.~Atiyah, V.~K.~Patodi and I.~M.~Singer,
  Math.\ Proc.\ Cambridge Phil.\ Soc.\  {\bf 77}, 43 (1975).
  doi:10.1017/S0305004100049410

\bibitem{Atiyah:1976jg} 
  M.~F.~Atiyah, V.~K.~Patodi and I.~M.~Singer,
  Math.\ Proc.\ Cambridge Phil.\ Soc.\  {\bf 78}, 405 (1976).
  doi:10.1017/S0305004100051872

\bibitem{Atiyah:1980jh} 
  M.~F.~Atiyah, V.~K.~Patodi and I.~M.~Singer,
  Math.\ Proc.\ Cambridge Phil.\ Soc.\  {\bf 79}, 71 (1976).
  doi:10.1017/S0305004100052105

\bibitem{Callan:1984sa} 
  C.~G.~Callan, Jr. and J.~A.~Harvey,
  Nucl.\ Phys.\ B {\bf 250}, 427 (1985).
  doi:10.1016/0550-3213(85)90489-4

\bibitem{Dai:1994kq} 
  X.~z.~Dai and D.~S.~Freed,
  J.\ Math.\ Phys.\  {\bf 35}, 5155 (1994)
  Erratum: [J.\ Math.\ Phys.\  {\bf 42}, 2343 (2001)]
  doi:10.1063/1.530747
  [hep-th/9405012].


\bibitem{Witten:2015aba} 
  E.~Witten,
  Rev.\ Mod.\ Phys.\  {\bf 88}, no. 3, 035001 (2016)
  doi:10.1103/RevModPhys.88.035001, 10.1103/RevModPhys.88.35001
  [arXiv:1508.04715 [cond-mat.mes-hall]].


\bibitem{Okumura:2016dsr} 
  K.~i.~Okumura and H.~Suzuki,
  arXiv:1608.02217 [hep-lat].

\bibitem{Makino:2016auf} 
  H.~Makino and O.~Morikawa,
  arXiv:1609.08376 [hep-lat].

\bibitem{Grabowska:2016bis} 
  D.~M.~Grabowska and D.~B.~Kaplan,
  arXiv:1610.02151 [hep-lat].


\end{thebibliography}
\end{document}